\documentclass{article}

\usepackage{cite}
\usepackage[cmex10]{amsmath}
\usepackage{algorithmic, algorithm}
\usepackage{array}
\usepackage{bm}
\usepackage[tight,footnotesize]{subfigure}
\usepackage{url}
\usepackage{bbm}
\usepackage{amsthm}
\usepackage{amsmath}
\usepackage{mathtools}
\usepackage{amssymb}
\usepackage{mathrsfs} 
\usepackage{graphicx}
\usepackage{color}
\usepackage{dsfont}
\usepackage[hang,flushmargin]{footmisc}
\usepackage{siunitx}

\newcommand{\cj}{\mathrm{j}}

\newcommand\mW{\mathbf W}
\newcommand\mB{\mathbf B}
\newcommand\mU{\mathbf U}

\newcommand\relu{\mathrm{ReLU}}

\newcommand\cn{\mathrm{cn}}
\newcommand\reals{\mathbb R}

\title{Deep Phase Decoder: Self-calibrating phase microscopy with an untrained deep neural network}

\author{Emrah Bostan$^{\ast}$, 
Reinhard Heckel$^{\dagger}$, 
Michael Chen$^{\ast}$, 
\\ Michael Kellman$^{\ast}$, and 
Laura Waller\thanks{Department of Electrical Engineering and Computer Sciences,
University of California, Berkeley, CA 94720, USA \newline
$^{\dagger}$ Department of Electrical and Computer Engineering, 
Technical University of Munich, Munich, Germany \newline
Corresponding author: emrah.bostan@gmail.com}
}
\date{}

\begin{document}
\maketitle

\begin{abstract}
Deep neural networks have emerged as effective tools for computational imaging including quantitative phase microscopy of transparent samples. To reconstruct phase from intensity, current approaches rely on supervised learning with training examples; consequently, their performance is sensitive to a match of training and imaging settings. Here we propose a new approach to phase microscopy by using an untrained deep neural network for measurement formation, encapsulating the image prior and imaging physics. Our approach does not require any training data and simultaneously reconstructs the sought phase and pupil-plane aberrations by fitting the weights of the network to the captured images. To demonstrate experimentally, we reconstruct quantitative phase from through-focus images blindly (i.e. no explicit knowledge of the aberrations).
\end{abstract}

\section*{Introduction}

\noindent Quantitative phase microscopy (QPM) enables label-free imaging of transparent samples such as unstained cells and tissues~\cite{Marquet.etal2005, Popescu.2011}, and non-absorbing micro-elements~\cite{Barty.etal1998}. QPM can use partially-coherent beams (in lieu of coherent ones~\cite{Rappaz.etal2014}) to increase spatial resolution and light throughput with reduced speckle. Examples include through-focus~\cite{Waller.etal2010, Descloux.etal2018}, interferometric~\cite{Wang.etal2011, Kim.etal2014}, and angle-scanning~\cite{Zheng.etal2013, Tian.Waller2015} microscopes. The common design theme is to image distinct non-linear renderings of phase as intensity from which quantitative phase is numerically recovered. For a given method, the performance and image quality is intrinsically governed by the phase reconstruction step.~\cite{Yeh.etal2015}.




Traditionally, the phase reconstruction problem is addressed by solving an inverse problem by minimizing least-squares loss that is based on the physics of the problem. The approach is fundamental to phase imaging~\cite{Fienup.1982} and has been practically employed in various systems. An immediate advantage is that our prior assumptions on the images can be directly integrated through regularization. An example is to constrain the phase image to admit a sparse representation in the wavelet domain~\cite{Pein.etal2016}. Such regularizers work well for various phase microscopes and improve the reconstruction quality~\cite{Kim.etal2014, Bostan.etal2016}. Another major advantage of the physics-based formulation is the possibility to incorporate algorithmic self-calibration~\cite{Ou.etal2014,Jingshan.etal2015}. It involves---in alternation with the phase retrieval step---minimizing the least-squares loss over unknown or partially-known system parameters such as pupil aberrations~\cite{Yeh.etal2015}. The concept hence accounts for the model-mismatch in the imaging pipeline. This provides us with great flexibility and allows phase reconstruction from measurements that are not fully characterized. In designing self-calibrating algorithms, 
the need for regularization (i.e., prior models for phase) is emphasized~\cite{Chen.etal18} since one simultaneously decouples the individual contributions of phase and aberrations to the measured images. However, typical regularization techniques are hand-crafted and require manual tuning of parameters even after the model is constructed.

\begin{figure*}[t!]
    \centering
    \includegraphics[scale=0.425]{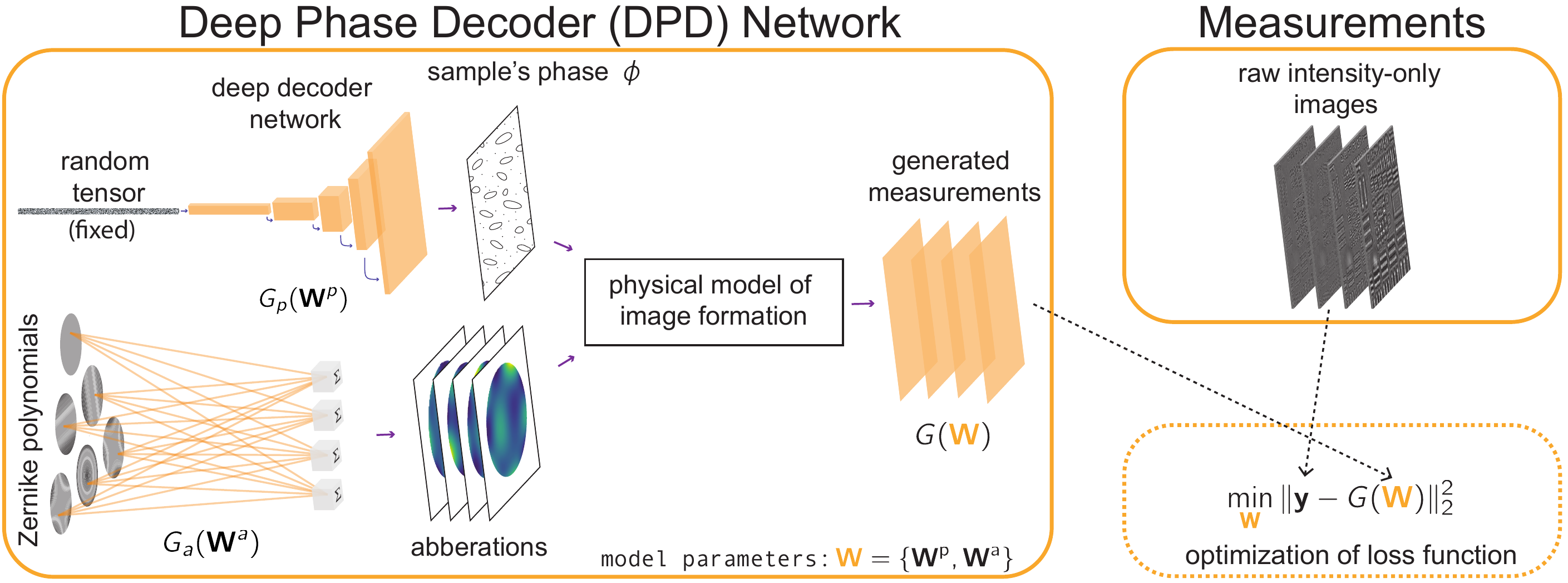}
    \caption{The deep phase decoder algorithm aims to minimize the Euclidean distance between the measured intensity images and the hypothetical ones generated by our untrained deep network. The optimization problem, which is nonlinear and nonconvex, is stated in terms of the network's weights and is solved iteratively using a gradient-based procedure. Once the weights are optimized, the sought phase image is retrieved as the output of the deep decoder part of the network.}
    \label{fig:workFlow}
\end{figure*}

More recently, deep neural networks (DNNs),
typically trained in an end-to-end fashion on large datasets to directly map given intensities back to phase, have been used to obtain efficient phase retrieval algorithms.
For phase microscopy, trained DNNs give state-of-the-art performance in holographic~\cite{Rivenson.etal2018}, lensless~\cite{Sinha.etal17}, ptychographic~\cite{Nguyen.etal18}, and through-scattering-media~\cite{Li.etal18,Borhani.etal18} configurations, among others~\cite{Jo.etal2019}. The results have validated the efficiency of properly trained DNNs to solve non-linear inverse problems and shifted the computational paradigm in QPM towards predominantly data-driven frameworks. 
However, for deep networks to work well, the proximity of training and experiment settings is critical as the performance is susceptible to variations in sample features, instrumentation, and acquisition parameters~\cite{Jo.etal2019}. Although improved DNN architectures have been proposed~\cite{Li.etal18b, Xue.etal2019, Kellman.etal2019, Kellman.etal2019b}, training-based approaches fundamentally rely on the reconstructed phase image to come from a distribution that is close to the one of the training images and are thus sensitive to misfits.

In this paper, we propose a new QPM algorithm that is based on a deep network, but requires no ground-truth training data.
Our approach is inspired by the idea of employing untrained generative DNNs as prior models for images, a concept that is pioneered by the so-called deep image prior~\cite{Ulyanov.etal2018}. 
Specifically, Ulyanvov et al.~\cite{Ulyanov.etal2018} fitted a noisy image via optimizing over the weights of a randomly initialized, over-parameterized autoencoder (i.e., an autoencoder with more weights than the number of image pixels), and observed that early stopping the regularization yields good denoising performance, an effect theoretically explained in~\cite{Heckel_Soltanolkotabi_2019}. 
For denoising, regularization through early stopping is critical, since the network can in principle fit the noisy data perfectly. 
Subsequently, an under-parameterized (i.e., less parameters than the number of image pixels) image-generating network, named the deep decoder, that does not need early stopping or any other further regularization has been proposed~\cite{Heckel.Hand2019}. The framework acts as a concise image model that provides a lower-dimensional description of images, akin to the sparse wavelet representations, and thus regularizes through its architecture alone. Unfortunately, a naive application of the method to the problem at hand would not account for practical issues such as drift and sample-induced aberrations~\cite{Chen.etal18}, which points to the need of properly incorporating our knowledge about optical physics to achieve self-calibration.


The key contribution of this paper is a DNN-based self-calibrating reconstruction algorithm for QPM that is training-free and recovers quantitative phase from raw images recorded without the explicit knowledge of aberrations. We specify the entire measurement formation as an untrained DNN whose weights are fitted to the recorded images. Leveraging the well-characterized system physics and non-linear forward model, our network combines a fully-connected layer that synthesizes aberrations from Zernike polynomials with the deep decoder that is used to generate phase. The proposed algorithm hence describes both the image and aberrations by a few weight coefficients, and as a consequence enables us to jointly retrieve the phase and individual aberration profile of each measurement without requiring any training data. We term our algorithm the deep phase decoder (DPD) and demonstrate it on a commercial widefield microscope.


\section*{Methods}

\begin{figure}[t!]
    \centering
    \includegraphics[scale=0.4175]{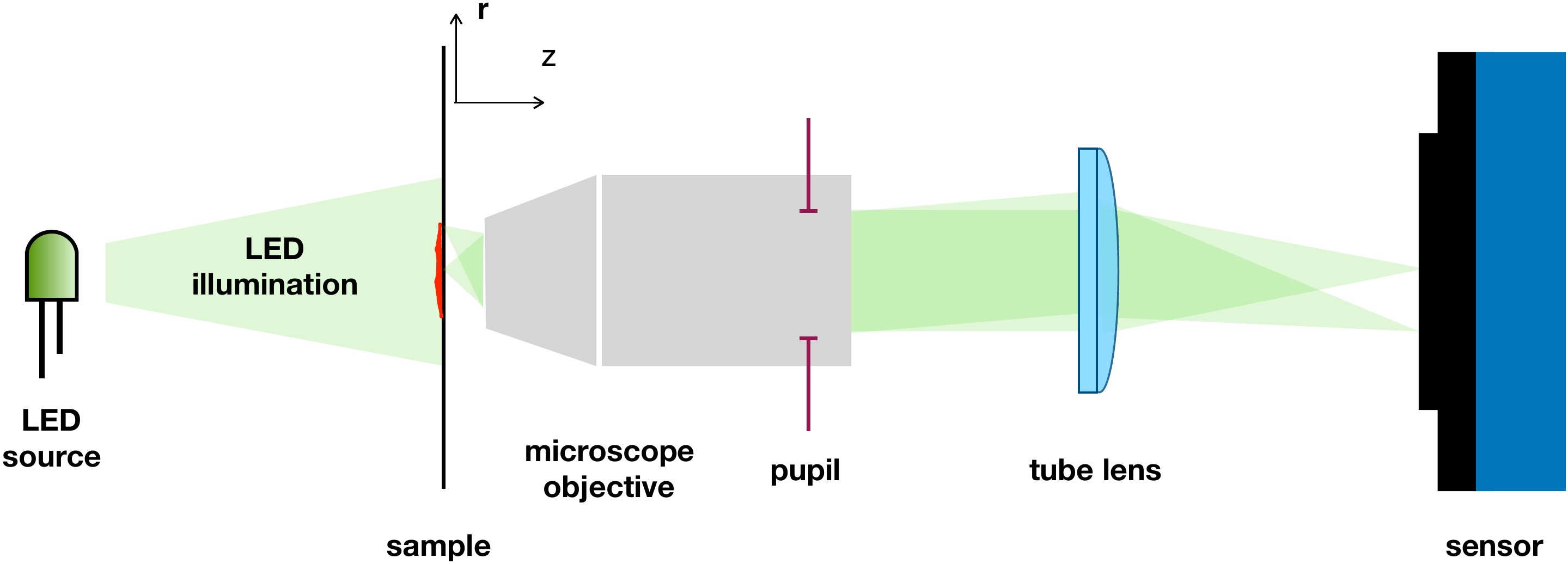}
    \caption{We experimentally validate our method with a dataset of intensity images captured in a microscope with the sample at varying defocus distances. The intensity images are then fed into the proposed algorithm to computationally reconstruct the sample's phase and wavefront aberrations without knowing the pupil functions (defocus distances) used during the data acquisition.}
    \label{fig:optical_setup}
\end{figure}

Next, we describe the image formation process in our optical setup (Fig.~\ref{fig:optical_setup}), and then describe our reconstruction approach in more detail. We consider an optically-thin and transparent sample that is placed at the focal plane of the microscope's objective. The sample's complex-valued image (i.e., its transmission function) is characterized as
\begin{equation}
    o(\mathbf{r}) = \exp \left( \cj \phi(\mathbf{r}) \right)\hspace{-0.23em},
    \label{eq:transmissionFunction}
\end{equation}
where $\phi$ represents the spatial distribution of phase over 2D coordinates $\mathbf{r}$. The LEDs are place sufficiently far away that their illumination can be modeled as a monochromatic plane wave at the sample plane. Thus, the irradiance of the beam impinging on the camera is given by
\begin{equation}
    y(\mathbf{r}) = \lvert c_{\rm psf} \ast o\lvert^2 (\mathbf{r})\text{,}
    \label{eq:forwardModelCont}
\end{equation}
where $\ast$ denotes spatial convolution and $c_{\rm psf}$ is the coherent point–spread function of the microscope. The sensor then measures the sampled irradiance, $\mathbf{y} \in \mathbb{R}^p$, where $p$ is the total number of pixels on the camera. In matrix form, 
\begin{equation}
    \mathbf{y} = \lvert \mathbf{F}^{-1} \mathbf{P}_{\rm circ} \mathbf{F} \mathbf{o} \lvert^2 \text{,}
    \label{eq:forwardModelDisc}
\end{equation}
where $\mathbf{F}$ is the discrete Fourier transform matrix and $\mathbf{P}_{\rm circ}$ is the ideal and space-invariant exit pupil function, which is a circle with its radius determined by numerical aperture (NA) of the objective and wavelength $\lambda$. 

Phase is recovered based on multiple images with some type of data diversity that translates phase information into intensity (e.g. defocus~\cite{Waller.etal2010}, illumination coding~\cite{Tian.etal2014}, pupil coding~\cite{Horisaki.etal2014}). Here we adopt a pupil-coding scheme where the wavefront at the exit pupil~\cite{Fienup.1982} is differently aberrated for each measurement. The pupil aberration is modeled as a weighted sum of Zernike polynomials, so it is parameterized by a small number of coefficients:
\begin{equation}
    \mathbf{P} = \mathbf{P}_{\rm circ}  \exp\left(  \cj \mathbf{Zc}\right) \text{,}
    \label{eq:pupilFunction}
\end{equation}
where the Zernike basis $\mathbf{Z}  = \left[ \mathbf{z}_1 \, \mathbf{z}_2 \, \dots \,\mathbf{z}_M \right]$ is composed of $M$ orthogonal modes in vectorized form and $\mathbf{c}$ contains the corresponding coefficients of each mode.  

The microscope is probed with a known (or pre-calibrated) set of aberrations $\left\lbrace\mathbf{P}\right\rbrace_{n=1}^N$, where $N$ is the total number of intensity images. The inverse problem then aims to recover the sample's transmission function as 
\begin{equation}
    \mathbf{o}^{\star} = \arg \underset{\mathbf{o}}{\min} \sum_{n=1}^N\| \sqrt{\mathbf{y}_n} - \lvert \mathbf{F}^{-1} \mathbf{P}_{n} \mathbf{F} \mathbf{o} \lvert \|_2^2 \text{.}
    \label{eq:phaseRetrieval}
\end{equation}
This can be solved by gradient-descent (or an accelerated variation), which is closely related to the well-known Gerchberg-Saxton method~\cite{Yeh.etal2015}. After solving for the complex-valued $\mathbf{o}^{\star}$, the phase image is its argument. The conventional phase recovery in~\eqref{eq:phaseRetrieval} does not necessarily impose any regularization on the recovered phase and the aberrations must be known a priori. To address these without needing any training data, we introduce a deep network in the derived formulation.

At the core of our approach, we use a DNN that generates $N$ intensity images. The network, denoted by $G(\mathbf{W})$, reparameterizes the measurement formation in~\eqref{eq:forwardModelDisc} in terms of a weight tensor $\mathbf{W}$ rather than pixels in complex-image space as in~\eqref{eq:phaseRetrieval}. The network is untrained and the weights, which are randomly initialized, are optimized by solving the following problem:
\begin{equation}
    \mathbf{W}^{\star} = \arg \underset{\mathbf{W}}{\min} \| \mathbf{Y} - G(\mathbf{W}) \|_2^2 \text{,}
    \label{eq:deepPhaseDecoder}
\end{equation} 
where $\mathbf{Y} = [ \sqrt{\mathbf{y}_1}  \dots \sqrt{\mathbf{y}_N} ] \in \mathbb{R}^{p \times N}$ accommodates all the measured data. Once the optimal weights $\mW^{\star}$ are obtained, phase is retrieved as the output of an appropriate layer in the network. The remarkable aspect---in terms of data requirement---is that the process is solely driven by the acquired images and does not involve any training data. The main reason is that the generative network's weights (and hence its output image) are adjusted on-the-fly in~\eqref{eq:deepPhaseDecoder} rather than training it to be able to represent a certain class of images.  

We design the network $G$ to encapsulate two sub-generators, $G_{\rm p}$ and $G_{\rm a}$, that synthesize a phase image and the pupil aberration of each individual measurement, respectively (see Fig.~\ref{fig:workFlow}). For the phase generating network $G_{\rm p}$, we use a deep decoder~\cite{Heckel.Hand2019}, which transforms a randomly chosen and fixed tensor $\mB_0 \in \reals^{n_0 \times k}$ consisting of $k$ many $n_0$-dimensional channels to an $n_d \times 1$ dimensional (i.e. gray-scale) image. In transforming the random tensor to a phase image, $G_p$ applies i) a pixel-wise linear combination of the channels, ii) upsampling, iii) rectified linear units (ReLUs), and iv) channel normalization. Specifically, the update at the $(i+1)$-th layer is given by
\[
\mB_{i+1} =  \cn(\relu( \mU_{i} \mB_i \mW^p_i ) ), \quad i = 0,\ldots, d-1.
\]
Here $\mW_i^p \in \reals^{k \times k}$ contains the coefficients for the linear combination of the channels and the operator $\mU_i \in \reals^{n_{i+1} \times n_i}$ performs bi-linear upsampling. This is followed by a channel normalization operation, $\cn(\cdot)$, which is equivalent to normalizing each channel individually to zero mean and unit variance, plus a bias term. A phase image, which is the output of the $d$-layer network, is then formed, with $\mW_{d}^p \in \reals^{k}$, as
\[
\boldsymbol{\phi} = 2 \pi \, \mathrm{sigmoid} (\mB_d \mW_{d}^p).
\]
The aberration-generating network, $G_{\rm a}(\mW^a)$, relies on the parameterization in~\eqref{eq:pupilFunction} represented as a fully-connected layer (i.e. linear combination of Zernike modes) and the matrix $\mW^a$ contains the Zernike coefficients for all measuruments.  In combining the outputs of $G_{\rm p}$ and $G_{\rm a}$ , we reproduce the physical image formation using~\eqref{eq:transmissionFunction},~\eqref{eq:forwardModelDisc}, and~\eqref{eq:pupilFunction} in the network's architecture. The framework is implemented in \textit{PyTorch}, allowing us to solve~\eqref{eq:deepPhaseDecoder} using gradient-based algorithms thanks to auto-differentation with respect to $\mW = \{ \mW^p, \mW^a \}$. Once the optimal weights $\mW^{\star}$ are obtained, the reconstructed phase is given by $G_p (\mW^{p\, \star})$ where $\mW^p = \{\mW_0^p,\ldots,\mW_{d}^p\}$.


We now explain some implicit aspects of the our method. First, we see from~\eqref{eq:deepPhaseDecoder} that $G(\mathbf{W}^{\star})$ replicates the recorded intensities as closely as possible in the least-squares sense. Therefore, regularization of phase is governed by the generative network's architecture for the images have to lie in its range. Specifically, both $G_p$ and $G_a$ under-parametrize their corresponding outputs (fewer weights than the number of pixels in generated images), so DPD imposes regularization on phase and aberrations. Moreover, once $G$ is constructed, the strength of regularization is not hand-tuned, as is typically done (such as adjusting the sparsity level for wavelet-based methods). It is also noteworthy that the DPD performs the phase reconstruction from randomly initialized (as $G$ is untrained) aberrations as opposed to other self-calibrating schemes that use theoretical pupils as initialization~\cite{Yeh.etal2015,Chen.etal18}.

\begin{figure}[t!]
    \centering
    \includegraphics[scale=0.46]{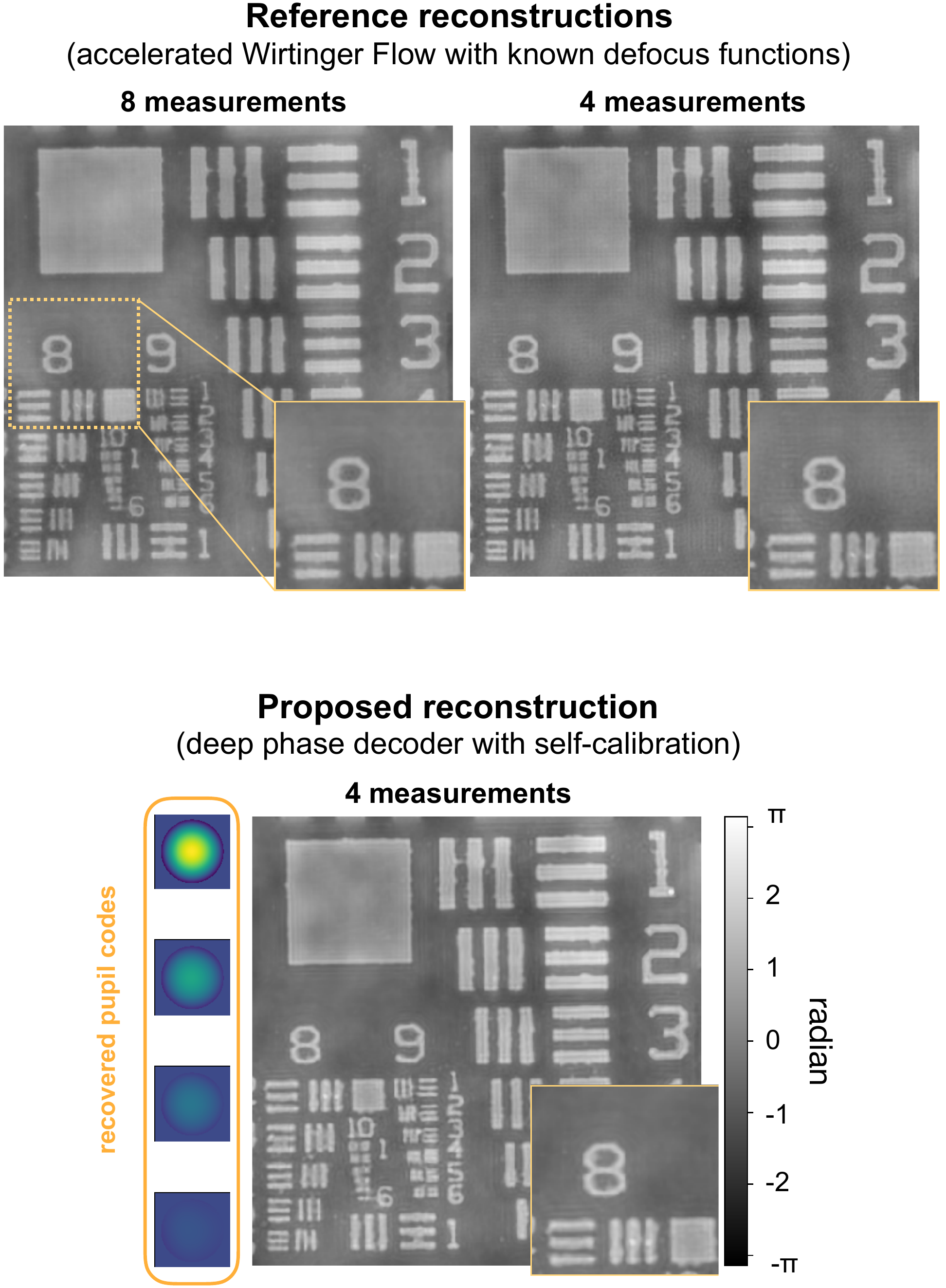}
    \caption{Experimental validation of the proposed reconstruction method from a stack of through-focus intensity images of a phase target with expected height of 150 nm (0.95 radians phase shift). (Left) Reconstructions by the accelerated Wirtinger flow algorithm~\cite{Candes.etal2015} are shown for comparison, with different numbers of measurements in the focus stack and known defocus distances. (Right) Our proposed algorithm's reconstruction achieves a similar phase result without any explicit knowledge of the aberrations.}
    \label{fig:results}
\end{figure}

\section*{Results}

To experimentally corroborate our method, we choose to use a commercial brightfield microscope (Nikon TE300) with LED illumination ($\lambda =\SI{0.514}{\micro\metre}$)~\cite{Tian.Waller2015}. A phase target (Benchmark Technologies) is imaged by a 40$\times $0.65 NA objective lens and intensity images are captured by a PCO.edge 5.5 sCMOS camera that is placed on the front port of the microscope, adding 2$\times$ magnification. To realize pupil-coding, we capture a through-focus stack of 8 images that are exponentially spaced (at $0, 1, 2, 4, 8, 16, 32$, and $\SI{64}{\micro\metre}$ defocus)~\cite{Jingshan.etal2014}. To compare against our method, we reconstruct reference phase images  with the accelerated Wirtinger flow algorithm~\cite{Candes.etal2015} using~\eqref{eq:phaseRetrieval} for 8 and 4 (defocus of $4, 8, 16$, and $\SI{32}{\micro\metre}$) measurements. We then use the same 4 measurements to solve the DPD optimization in~\eqref{eq:deepPhaseDecoder} using the RMSProp algorithm 
with $5\times 10^4$ iterations. The network is constructed with the following parameters: $k=32$, $n_0=16\times16$, and $n_d=512\times512$. Bi-linear upsampling is fixed to a factor of 2, making $G_p$ a 6-layer network. We use the first 9 Zernike polynomials after piston for $G_a$. The reconstructions from both methods show good agreement with each other (Fig.~\ref{fig:results}). Also, DPD jointly recovers defocus-like pupil functions, as expected. This experimentally validates our algorithm's ability to blindly reconstruct a reliable phase image from the measured intensities.


\section*{Conclusion}
In summary, we derived a new phase imaging algorithm that uses an untrained neural network, and demonstrate it on a phase-from-defocus dataset. Our DPD method, unlike its deep learning counterparts that are supervised, is training-free and does not rely on closely-matching training and experiment conditions. Moreover, our method is self-calibrating, allowing us to directly reconstruct high quality phase without a priori knowledge of the system's aberrations. 

\section*{Acknowledgments}
The authors thank Gautam Gunjala for help with Zernike polynomials. This work was supported by STROBE: A National Science Foundation Science and Technology Center under Grant No. DMR 1548924, and ONR grant. E. Bostan is supported by the Swiss National Science Foundation (SNSF) under grant P2ELP2 172278. M. Kellman is supported in part by the National Science Foundation (NSF) under grant DGE 1106400.

\bibliographystyle{IEEEbib}
\bibliography{sample}

\end{document}